\documentclass{aa}  

\usepackage{graphicx}
\usepackage{txfonts}
\begin{document}

   \title{Hidden or missing outflows in highly obscured galaxy nuclei?\thanks{{\it Herschel} is an ESA space observatory with science instruments provided by European-led Principal Investigator consortia and with important participation from NASA.}}

   \author{N.~Falstad\inst{1}
     \and
     F.~Hallqvist\inst{1}
     \and
     S.~Aalto\inst{1}
     \and
     S.~K{\"o}nig\inst{1}
     \and
     S.~Muller\inst{1}
     \and
     R.~Aladro\inst{2}
     \and
     F.~Combes\inst{3}
     \and
     A.~S.~Evans\inst{4,5}
     \and
     G.~A.~Fuller\inst{6}
     \and
     J.~S.~Gallagher\inst{7}
     \and
     S.~Garc{\'i}a-Burillo\inst{8}
     \and
     E.~Gonz{\'a}lez-Alfonso\inst{9}
     \and
     T.~R.~Greve\inst{10,11}
     \and
     C.~Henkel\inst{2,12}
     \and
     M.~Imanishi\inst{13}
     \and
     T.~Izumi\inst{13}
     \and
     J.~G.~Mangum\inst{4}
     \and
     S.~Mart{\'i}n\inst{14,15}
     \and
     G.~C.~Privon\inst{16}
     \and
     K.~Sakamoto\inst{17}
     \and
     S.~Veilleux\inst{18}
     \and
     P.~P.~van~der~Werf\inst{19}
          }

   \institute{Department of Space, Earth and Environment, Chalmers University of Technology, Onsala Space Observatory,
     439 92 Onsala, Sweden \\
     \email{niklas.falstad@chalmers.se}
     \and
     Max-Planck-Institut f{\"u}r Radioastronomie, Auf dem H\"ugel 69, 53121, Bonn, Germany
     \and
     Observatoire de Paris, LERMA, College de France, CNRS, PSL Univ., UPMC, Sorbonne Univ., F-75014, Paris, France
     \and
     National Radio Astronomy Observatory, 520 Edgemont Road, Charlottesville, VA 22903, USA
     \and
     Department of Astronomy, 530 McCormick Road, University of Virginia, Charlottesville, VA 22904, USA
     \and
     Jodrell Bank Centre for Astrophysics, School of Physics \& Astronomy, The University of Manchester, Oxford Road, Manchester M13\,9PL, UK
     \and
     Department of Astronomy, University of Wisconsin-Madison, 5534 Sterling, 475 North Charter Street, Madison WI 53706, USA
     \and
     Observatorio de Madrid, OAN-IGN, Alfonso XII, 3, E-28014-Madrid, Spain
     \and
     Universidad de Alcal{\'a}, Departamento de F{\'i}sica y Matem{\'a}ticas, Campus Universitario, E-28871 Alcal{\'a} de Henares, Madrid, Spain
     \and
     Cosmic Dawn Center (DAWN), DTU-Space, Technical University of Denmark, Elektrovej 327, DK-2800 Kgs. Lyngby; Niels Bohr Institute, University of Copenhagen, Juliane Maries Vej 30, DK-2100 Copenhagen $\O$
     \and
     Department of Physics and Astronomy, University College London, Gower Street, London WC1E\,6BT, UK
     \and
     Astron. Dept., King Abdulaziz University, P.O. Box 80203, 21589 Jeddah, Saudi Arabia
     \and
     National Astronomical Observatory of Japan, National Institutes of Natural Sciences (NINS), 2-21-1 Osawa, Mitaka, Tokyo 181–8588, Japan
     \and
     European Southern Observatory, Alonso de C{\'o}rdova 3107, Vitacura  763 0355, Santiago, Chile
     \and
     Joint ALMA Observatory, Alonso de C{\'o}rdova 3107, Vitacura 763 0355, Santiago, Chile
     \and
     Department of Astronomy, University of Florida, 211 Bryant Space Sciences Center, Gainesville, 32611 FL, USA
     \and
     Institute of Astronomy and Astrophysics, Academia Sinica, PO Box 23-141, 10617 Taipei, Taiwan
     \and
     Department of Astronomy, University of Maryland, College Park, MD 20742, USA
     \and
     Leiden Observatory, Leiden University, P.O.\ Box 9513, NL-2300 RA Leiden, The Netherlands
   }

   \date{}

 
  \abstract
      {Understanding the nuclear growth and feedback processes in galaxies requires investigating their often obscured central regions. One way to do this is to use (sub)millimeter line emission from vibrationally excited HCN (HCN-vib), which is thought to trace warm and highly enshrouded galaxy nuclei. It has been suggested that the most intense HCN-vib emission from a galaxy is connected to a phase of nuclear growth that occurs before the nuclear feedback processes have been fully developed.}
      {We aim to investigate if there is a connection between the presence of strong HCN-vib emission and the development of feedback in (ultra) luminous infrared galaxies ((U)LIRGs).}
   {We collected literature and archival data in order to compare the luminosities of rotational lines of HCN-vib, normalized to the total infrared luminosity, to the median velocities of $119$~$\mu$m OH absorption lines, potentially indicating outflows, in a total of 17 (U)LIRGs.} 
   {The most HCN-vib luminous systems all lack signatures of significant molecular outflows in the far-infrared OH absorption lines. However, at least some of the systems with bright HCN-vib emission do have fast and collimated outflows that can be seen in spectral lines at longer wavelengths, including in millimeter emission lines of CO and HCN (in its vibrational ground state) as well as in radio absorption lines of OH.}
   {We conclude that the galaxy nuclei with the highest $L_{\mathrm{HCN-vib}}/L_{\mathrm{IR}}$ do not drive wide-angle outflows detectable using the median velocities of far-infrared OH absorption lines. It is possible that this is due to an orientation effect where sources which are oriented in such a way that their outflows are not along our line of sight also radiate a smaller proportion of their infrared luminosity in our direction. It could also be that massive wide-angle outflows destroy the deeply embedded regions responsible for bright HCN-vib emission, so that the two phenomena cannot coexist. This would strengthen the idea that vibrationally excited HCN traces a heavily obscured stage of evolution before nuclear feedback mechanisms are fully developed.}


   \maketitle
%

   \section{Introduction}
   Luminous and ultraluminous infrared galaxies ((U)LIRGs) are gas-rich systems that radiate intensely in the infrared portion of the electromagnetic spectrum, with infrared luminosities ($L_{\mathrm{IR}}=L(8\text{--}1000$~$\mathrm{\mu m})$) in excess of $10^{11}L_{\sun}$ and $10^{12}L_{\sun}$, respectively \citep[e.g.,][]{san96}. These large luminosities are due to dust-reprocessed radiation from intense star formation, an active galactic nucleus (AGN), or both. In the local Universe, (U)LIRGs are relatively rare compared to less luminous objects, but surveys at millimeter and submillimeter wavelengths have shown that they are much more numerous at high redshifts \citep[e.g.,][]{sma97,hug98}, indicating that they play an important role in the evolution of galaxies. Two important questions in this context are: how did the supermassive black holes (SMBHs) found in the nuclei of most galaxies grow, and how are they related to the evolution of their host galaxies? From observations, we know that tight relations between the mass of the central SMBH and various properties of the host galaxy exist in large ellipticals \citep[e.g.,][]{mag98,fer00,geb00,kor13}. It has been suggested that these relations were partly established through a process, mostly taking place in the early Universe, in which galaxies of similar masses collide and merge \citep[see, e.g.,][for a review]{kor13}. In such mergers, large amounts of gas are funneled into the new nucleus, giving rise to intense starbursts, AGN activity, or both. The exact physics of how the black hole-host relations would be established are still obscure, but several suggested galaxy merger scenarios include a starburst dominated phase followed by a phase of obscured AGN activity. Eventually, the AGN becomes strong enough to drive outflows that clear the surrounding material and reveal the central activity \citep[e.g.,][]{san88a,hop06,urr08,sim12}. Thus, studies of the most dust-embedded phase of evolution and the early stages of feedback are essential in order to understand how SMBHs grow together with their host galaxies.

   However, observations of the most enshrouded galaxy nuclei are hampered by the large amounts of obscuring material that surround them. One solution to this problem is to observe at millimeter and radio wavelengths, where the dust is less optically thick \citep[e.g.,][]{bar15}. Some common tracers of obscured galaxy nuclei at these wavelengths are the low-$J$ rotational lines of the HCN and HCO$^{+}$ molecules. Recently, however, it has been shown that, in the most obscured systems, even these dense-gas tracers are heavily affected by continuum- and/or self-absorption by cooler foreground gas \citep[e.g.,][]{sak09,aal15b,mar16}. A better suited tracer of the most obscured systems is offered by HCN in its first vibrationally excited state, $v_{2}=1$ (hereafter HCN-vib). Strong emission lines from rotational transitions within the $v_{2}=1$ state have been detected in external galaxies at (sub)millimeter wavelengths, first by \citet{sak10} and then in many subsequent observations. With energy levels that lie more than $1000$~K above the ground state, HCN-vib is mainly excited by absorption of mid-infrared radiation. However, to be efficiently populated it requires brightness temperatures in excess of $100$~K at $14$~$\mu$m, the wavelength of the transition between the ground and first vibrationally excited states of HCN, translating into a H$_{2}$ column density larger than $2\times10^{23}$~cm$^{-2}$ for a dust temperature of $100$~K \citep{aal15b}. This makes bright HCN-vib emission an excellent unabsorbed \citep{mar16} tracer of high column density gas with high mid-infrared surface brightness. While the direct mid-infrared radiation from such regions is often hidden from us by the large quantities of obscuring material, the rotational transitions inside the vibrationally excited states occur at millimeter and submillimeter wavelengths where the dust opacity is lower. The low-J rotational lines of HCN-vib are thus useful probes of the nuclear mid-infrared source in heavily obscured objects. Since the first extragalactic (sub)millimeter detection of HCN-vib rotational line emission by \citet{sak10}, the field has developed rapidly \citep[e.g.,][]{ima13,aal15a,aal15b,ima16a,ima16b,ima18,mar16}, notably thanks to the Atacama Large Millimeter/submillimeter Array (ALMA).

   In their study of HCN-vib in obscured galaxies, \citet{aal15b} found that all galaxies with detected HCN-vib emission show evidence of inflows, outflows, or both. In addition, they found a tentative trend that galaxies with fast molecular outflows have fainter HCN-vib emission relative to their total infrared luminosity. The trend is especially striking when comparing the strength of the HCN-vib emission to the velocity of the wide-angle OH outflows studied by, for example, \citet{vei13}. Based on a relatively small sample of nine galaxies, \citet{aal15b} tentatively suggest that strong HCN-vib emission is connected to a rapid phase of nuclear growth that occurs right before the onset of strong feedback.

   In this paper, we present the results of a study employing a sample of $19$ galaxies including data from the literature as well as previously unpublished data, in order to further investigate the connection, or lack thereof, between strong HCN-vib emission and molecular outflows. The sample used in this study is described in Sect. \ref{sec:sample} and in Sect. \ref{sec:results} we present the results which are then discussed in Sect. \ref{sec:discussion}. Finally, the conclusions that we draw from the analysis are summarized in Sect. \ref{sec:conclusions}. Throughout the paper, a $H_{0} =73$ km\,s$^{-1}$\,Mpc$^{-1}$, $\Omega_{\mathrm{m}} = 0.27$, and $\Omega_{\Lambda} = 0.73$ cosmology is adopted.
   
   \section{Sample}\label{sec:sample}
   We have searched for (U)LIRGs with existing ALMA observations of HCN-vib that are either already published or that have the data publicly available in the the ALMA science archive\footnote{The ALMA science archive is available at https://almascience.nrao.edu/alma-data/archive.}. In addition, we have included published HCN-vib observations taken with the Plateau de Bure Interferometer (PdBI), the Northern Extended Millimeter Array (NOEMA), or the Submillimeter Array (SMA). We have not included non-detections, unless they have existing observations of the OH doublet at $119$~$\mu$m, which we use to determine the outflow velocity. Non-detections have also been excluded when a meaningful upper limit to the HCN-vib flux could not be determined, either due to low sensitivity spectra or due to significant flux from the HCO$^{+}$ line which is separated from HCN-vib by ${\sim}400$~km\,s$^{-1}$. This is a possible source of bias, as HCN-vib emission may be hidden by the HCO$^{+}$ line in sources with outflowing gas. In practice, most sources that have been excluded either do not have any outflow signatures in the far-infrared OH lines that we use to trace outflows (see below), or have not been observed in these lines. The two exceptions are IRAS~19254-7245 \citep{ima16a} and NGC 6240 (ALMA project 2015.1.01448.S, PI: R. Tunnard), in which the median velocities of the OH $119$~$\mu$m lines are ${\sim}-250$ and ${\sim}-200$~km\,s$^{-1}$, respectively. In total, $10$ sources have been excluded, most of them due to lack of OH measurements.

   For consistency of the sample, we ensure that outflow velocities were measured using the same common tracers and following the same procedures. All sources with HCN-vib observations have therefore been checked for observations of the far-infrared OH doublet at $119$~$\mu$m with the \emph{Herschel} Space Observatory \citep{pil10} as this is a known tracer of molecular outflows on subkiloparsec scales, which has been observed in a large number of (U)LIRGs \citep[e.g.,][]{fis10,vei13,gon17}. The OH molecule has high abundances in photodissociation regions as well as in X-ray dominated regions \citep[e.g.,][]{mal96,goi11,mei11}. Unlike other common outflow tracers like the millimeter rotational transitions of CO and HCN, the ground state OH $119$~$\mu$m doublet is often detected in absorption towards the far-infrared background \citep[e.g.,][]{vei13}, providing an unambiguous tracer of the gas motion. The transitions between excited states (for example the doublets at $65$ and $84$~$\mu$m) are generally weaker and the line wings are not always detected \citep{gon17}. Recently, the $119$~$\mu$m doublet was also used to detect a fast outflow ($\lesssim800$~km\,s$^{-1}$) in a galaxy at a redshift of $5.3$ \citep{spi18}. Systematic searches of outflow signatures in the far-infrared OH lines have revealed evidence of molecular outflows in approximately two thirds of the observed (U)LIRG samples \citep{vei13,spo13}. From this detection rate, \citet{vei13} inferred a wide average opening angle of ${\sim}145\degr$, assuming that all objects in their sample have outflows. In fact, based on radiative transfer models as well as spatially resolved observations of two outflows, \citet{gon17} argue that the $119$~$\mu$m OH absorptions are primarily sensitive to wide-angle outflows. On the other hand, unambiguous signatures of inflows are only found in one tenth of the sources in the sample of \citet{vei13}, suggesting that OH inflows are either rare or have, on average, smaller opening angles.
     
In particular, the 43 sources in the far-infrared OH sample of \citet{vei13} have been checked for observations of the (sub)millimeter HCN-vib $J=3\text{--}2$ or $J=4\text{--}3$ lines. Fourteen of these had HCN-vib detections or data that allowed for meaningful upper limits to be estimated. One of them, IRAS~05189-2524 (see Appendix \ref{app:HCN}), has previously unpublished data in the ALMA science archive while data for the remaining thirteen were taken from \citet{sak10,aal15a,aal15b,ima16a,ima16b,mar16,pri17,ala18}; and K{\"o}nig et al.~(in prep.). In addition, five sources not included in the sample of \citet{vei13} were found to have usable observations of HCN-vib. One, IRAS 12224-0624 (see Appendix \ref{app:HCN}), has previously unpublished data in the ALMA science archive while the other four were taken from \citet{aal15b} and \citet{ima16a}. Three of these five sources, Zw~049.057, IRAS~20414-1651, and NGC 7469, also have existing observations of the OH $119$~$\mu$m doublet (see Appendix \ref{app:OH}).

In total, 19 (U)LIRGs with existing HCN-vib observations were found. Of these, 17 also have observations of the OH $119$~$\mu$m doublet. The final sample, with the adopted redshift and distance of each source, is presented in Table \ref{tab:sample}.

\begin{table*}
  \caption{Sample galaxies.}\label{tab:sample}
  \centering
  \begin{tabular}{lcccc}
    \hline
    \hline
    Name & Type\tablefootmark{a} & $z$\tablefootmark{b} & $D_{L}$\tablefootmark{c} & $L_{\mathrm{IR}}$\tablefootmark{d} \\
    & & & (Mpc) & ($10^{11}$ L$_{\sun}$) \\
    \hline
    NGC 4418 & Sy 2 & $0.0071$ & $33.6\pm2.5$ & $1.32\pm0.19$ \\ 
    IC 860 & \ion{H}{ii} & $0.0130$ & $62.1\pm4.3$ & $1.60\pm0.22$ \\	  
    IRAS 12224-0624 & L & $0.0264$ & $119.5\pm8.2$ & $1.98\pm0.27$ \\ 
    Zw 049.057 & \ion{H}{ii} & $0.0130$ & $62.4\pm4.3$ & $2.04\pm0.28$ \\  
    NGC 7469 & Sy 1 & $0.0164$ & $68.2\pm4.7$ & $4.13\pm0.57$ \\
    I Zw 1 & Sy 1 & $0.0610$ & $258.2\pm17.7$ & $8.22\pm1.13$ \\ 
    UGC 5101 & L &$0.0394$ & $169.7\pm11.6$ & $9.50\pm1.30$ \\
    IRAS 20551-4250 & \ion{H}{ii} & $0.0430$ & $185.2\pm12.7$ & $10.65\pm1.46$ \\ 
    IRAS 15250+3609 & L & $0.0552$ & $243.7\pm16.7$ & $11.08\pm1.53$ \\ 
    IRAS 08572+3915 & L & $0.0584$ & $253.1\pm17.3$ & $13.38\pm1.84$ \\ 
    IRAS 05189-2524 & Sy 2 & $0.0428$ &  $180.2\pm12.4$ & $13.50\pm1.85$ \\
    IRAS 22491-1808 & \ion{H}{ii} & $0.0778$ & $337.0\pm23.1$ & $13.88\pm1.92$ \\ 
    Mrk 273 & Sy 2 &$0.0378$ & $165.6\pm11.3$ & $14.81\pm2.03$ \\
    IRAS 20414-1651 & \ion{H}{ii} & $0.0871$ & $383.4\pm26.3$ & $15.12\pm2.18$ \\
    Arp 220 & L & $0.0181$ & $84.1\pm5.8$ & $17.25\pm2.36$ \\ 
    IRAS 13120-5453 & Sy 2 & $0.0308$ & $137.8\pm9.5$ & $19.20\pm2.65$ \\ 
    IRAS 12112+0305 & L & $0.0733$ & $326.3\pm22.4$ & $20.35\pm2.79$ \\  
    IRAS 17208-0014 & \ion{H}{ii} & $0.0428$ & $189.3\pm13.0$ & $26.21\pm3.59$ \\
    Mrk 231 & Sy 1 &$0.0422$ & $184.3\pm12.6$ & $34.30\pm4.70$ \\ 
    \hline
  \end{tabular}
  \tablefoot{
    \tablefoottext{a}{Optical spectral types, \ion{H}{ii}, L (for LINER), Sy 1 and Sy 2 (for Seyfert 1 and 2, respectively). When available, we have used the classifications from \citet{vei95} and \citet{kim98b}, otherwise, we have used classifications from \citet{san88b}, \citet{baa98}, and \citet{ver01}.}
  \tablefoottext{b}{In priority order, we have used redshifts from \citet{vei13}, \citet{san03}, or \citet{stra92}, except for IRAS 05189-2524 where the redshift from \citet{san91} was used.}
    \tablefoottext{c}{Luminosity distances were calculated from the redshifts following the same procedure as \citet{san03}.}
    \tablefoottext{d}{Infrared luminosities were calculated using the prescription in Table 1 of \citet{san96} (originally from \citet{per87}) using IRAS fluxes taken from \citet{san03}, except for IRAS~20414-1651 where fluxes were taken from \citet{mos92}, and I~Zw~1 where they were taken from \citet{san89}.}
} 
\end{table*}

\section{Results}\label{sec:results}
Vibrationally excited HCN luminosities, $L_{\mathrm{HCN-vib}}/L_{\mathrm{IR}}$ ratios, and OH median velocities are presented in Table \ref{tab:properties}. We adopt the $L_{\mathrm{HCN-vib}}/L_{\mathrm{IR}}$ ratio as a parameter to describe the strength of HCN-vib in a normalized fashion. However, we are aware that some part of the total $L_{\mathrm{IR}}$ may be unrelated to the dusty nucleus, and instead come from, for example, an extended starburst. In particular, in systems with multiple nuclei, for example Arp~220 and IRAS~12112+0305, $L_{\mathrm{IR}}$ will have contributions from both nuclei. In our sample, we are not aware of any sources were these effects would affect the $L_{\mathrm{HCN-vib}}/L_{\mathrm{IR}}$ ratio enough to significantly change our results. We use the median velocity of the OH lines instead of the terminal outflow velocity as the former provides more robust values \citep{vei13}. Velocities are only given for those sources with absorption in the OH $119$~$\mu$m doublet, as this means that the gas is in the foreground and that a positive or negative velocity shift can be interpreted as evidence of gas moving towards or away from the background nucleus, respectively. We note that this does not necessarily indicate in- or outflowing gas in all cases as, for example, interactions between two nuclei can also affect the OH kinematics. In addition, it is important to take the uncertainties in the velocity determination into account. For example, \citet{vei13} define an outflow as having an OH absorption feature with a median velocity more negative than $-50$~km\,s$^{-1}$, which is the typical uncertainty on the velocities. The requirement to detect the OH doublet in absorption is a possible source of bias as this requires a high enough column density of OH in front of a strong enough far-infrared background source. However, we note that all galaxies with an HCN-vib detection, if observed, have also been detected in absorption in the OH $119$~$\mu$m doublet. Therefore, no sources with a combination of high HCN-vib luminosity and a fast outflow have been missed due to this effect.

The line luminosities presented in Table \ref{tab:properties} have been calculated following Eq. (1) in \citet{sol05}, applied to HCN-vib:
\begin{equation}
L_{\mathrm{HCN-vib}}=1.04\times10^{-3}S_{\mathrm{HCN-vib}}\,\Delta v\,\nu_{\mathrm{rest}}(1+z)^{-1}D_{\mathrm{L}}^{2},
\end{equation}
where $L_{\mathrm{HCN-vib}}$ is the HCN-vib luminosity measured in $L_{\odot}$, $S_{\mathrm{HCN-vib}}\,\Delta v$ is the velocity integrated flux in Jy\,km\,s$^{-1}$, $\nu_{\mathrm{rest}}$ is the rest frequency in GHz, and $D_{\mathrm{L}}$ is the luminosity distance in Mpc. Due to differences in the adopted redshifts and distances, these values differ slightly from those given in the original reference for some of the sources. In case of non-detections, we have included the $3\sigma$ upper limit to the line luminosity. In the case of IRAS 15250+3609, the HCN-vib line is likely blended with a potential outflow signature in the nearby HCO$^{+}$ line. Here, we have attributed the entire emission feature to the HCN-vib line, and the value given should be considered an upper limit to the HCN-vib luminosity. A short discussion of this source as well as descriptions of the analysis of previously unpublished HCN-vib observations and new OH outflow measurements can be found in Appendices \ref{app:15250}, \ref{app:HCN}, and \ref{app:OH}.

\begin{table*}
  \caption{$L_{\mathrm{HCN-vib}}$ and outflow properties.}\label{tab:properties}
  \centering
  \begin{tabular}{lccccc}
    \hline
    \hline
    Name & $L_{\mathrm{HCN-vib}}$ & $L_{\mathrm{HCN-vib}}/L_{\mathrm{IR}}$ & $v_{\mathrm{50}}$(abs)\tablefootmark{a} & Observatory & References \\
    & ($10^{3} L_{\sun}$) & ($10^{-8}$) & (km\,s$^{-1}$) &  & \\
    \hline
    $J=3\text{--}2$ & & &  & &  \\
    NGC 4418 & $5.0\pm1.2$ & $3.77\pm0.71$ & $111$ & SMA & 1 \\
    IC 860 & $4.2\pm0.8$ & $2.65\pm0.37$ & not obs. & PdBI & 2  \\
    Zw 049.057 & $7.3\pm2.0$ & $3.57\pm0.76$ & $37$ & PdBI & 2 \\
    NGC 7469 & $<0.1$ & $<0.03$ & emi. & ALMA & 3  \\
    I Zw 1 & $<1.7$ & $<0.21$ & emi. & ALMA & 3  \\
    UGC 5101 & $20.5\pm5.5$ & $2.15\pm0.50$ & $-9$ & NOEMA & 4 \\
    IRAS 20551-4250 & $2.3\pm0.7$ & $0.21\pm0.06$ & $-381$ & ALMA & 5 \\
    IRAS 15250+3609 & $<11.3$\tablefootmark{b} & $<1.02$ & $189$ & ALMA & 3 \\
    IRAS 08572+3915 NW & $<6.4$ & $<0.48$ & $-489$ & ALMA & 3 \\
    IRAS 22491-1808 E & $13.2\pm3.9$ & $0.95\pm0.25$ & $99$ & ALMA & 3 \\
    Mrk 273 & $<16.9$ & $<1.14$ & $-201$ & NOEMA & 6 \\
    IRAS 20414-1651 & $<7.5$ & $<0.50$ & $-32$ & ALMA & 3 \\  
    Arp 220 W & $40.9\pm7.0$ & $2.37\pm0.25$\tablefootmark{c} & $21$ & ALMA & 7 \\
    Arp 220 E & $8.9\pm1.5$ & $0.51\pm0.05$\tablefootmark{c} & $21$ & ALMA & 7 \\
    IRAS 12112+0305 NE & $14.3\pm4.0$ & $0.70\pm0.17$\tablefootmark{c} & $-117$ & ALMA & 3 \\ 
    Mrk 231 & $14.5\pm3.1$ & $0.42\pm0.07$ & $-237$ & PdBI & 8 \\
    \hline
    $J=4\text{--}3$ & & & &  \\

    NGC 4418 & $12.4\pm2.5$ & $9.41\pm1.25$ & $111$ & SMA & 1\\
    IRAS 12224-0624 & $16.9\pm4.5$ & $8.52\pm1.95$ & not obs. & ALMA &  9 \\
    IRAS 20551-4250 & $4.8\pm1.2$ & $0.45\pm0.09$ & $-381$ & ALMA & 10 \\
    IRAS 05189-2524 & $8.0\pm2.7$ & $0.59\pm0.18$ & $-387\tablefootmark{d}$ & ALMA & 9 \\
    Arp 220 W & $108.1\pm19.9$ & $6.27\pm0.77$ & $21$ & ALMA & 7 \\
    Arp 220 E & $25.8\pm4.4$ & $1.49\pm0.15$ & $21$ & ALMA & 7 \\
    IRAS 13120-5453 & $<5.5$ & $<0.29$ & $-195$ & ALMA & 11 \\
    IRAS 17208-0014 & $101.9\pm17.5$ & $3.89\pm0.40$ & $51$ & ALMA & 2 \\
    \hline
  \end{tabular}
  \tablefoot{
    \tablefoottext{a}{Median OH outflow velocities taken from \citet{vei13} or, for sources not in their sample, from archival data following the same procedure. Sources that have not been observed or where the doublet was detected in emission are indicated with not obs. or emi., respectively. The typical uncertainty on the velocities is $50$~km\,s$^{-1}$.}
    \tablefoottext{b}{\citet{ima16a} interpret an emission feature at the position of the HCN-vib line as a sub-peak of the nearby HCO$^{+}$ line. Here, we use this value as an upper limit to the HCN-vib luminosity.}
    \tablefoottext{c}{The $L_{\mathrm{HCN-vib}}/L_{\mathrm{IR}}$ is calculated using the total $L_{\mathrm{IR}}$ of the system, possibly introducing some bias in systems with multiple nuclei.}   
    \tablefoottext{d}{The median velocity of the lines in IRAS~05189-2524 has been adjusted to correct for the different redshift adopted compared to the one used by \citet{vei13}.}
  }
  \tablebib{
 (1)~\citet{sak10}; (2)~\citet{aal15b}; (3)~\citet{ima16a}; (4) K{\"o}nig et al.~(in prep.); (5)~\citet{ima16b}; (6)~\citet{ala18}; (7)~\citet{mar16}; (8)~\citet{aal15a}; (9)~This work; (10)~\citet{ima13};  (11)~\citet{pri17}
     }
\end{table*}

\begin{figure*}
   \centering
   \includegraphics{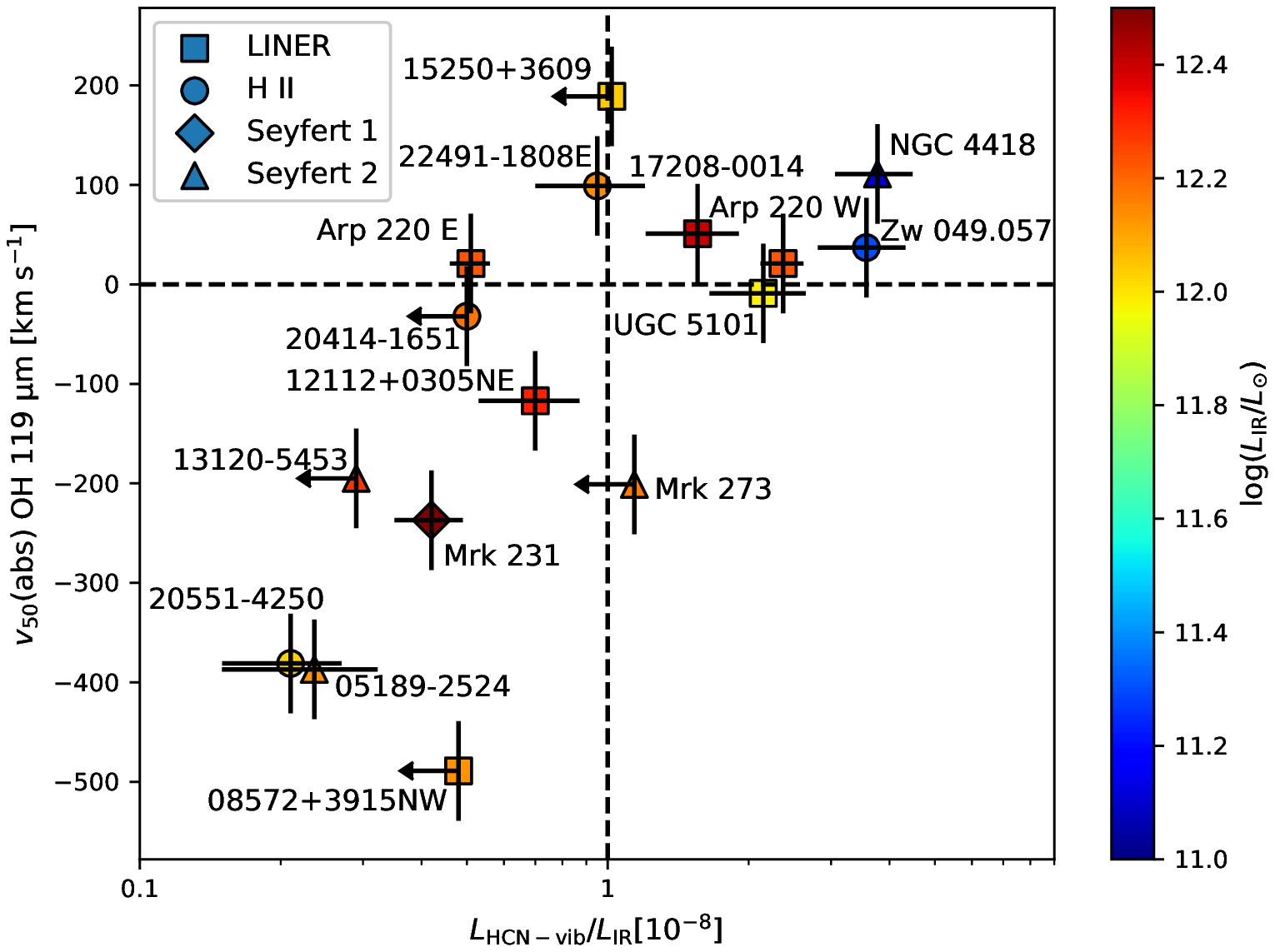}
   \caption{Median OH absorption velocity as a function of the HCN-vib luminosity relative to the the total infrared luminosity. For sources with observations of the $J=3\text{--}2$ transition, only this line is plotted. For sources with detections only in the $J=4\text{--}3$ line, the HCN-vib luminosity has been scaled down by a factor of $2.5$, with an extra uncertainty of $20\%$ included in the error bars. Squares, circles, diamonds, and triangles represent LINER, \ion{H}{ii}, Seyfert 1, and Seyfert 2 optical spectral types, respectively. Colors indicate the total infrared luminosity of each system. Upper limits ($3\sigma$) are indicated with arrows.}
              \label{fig:HCN_vs_OH}%
    \end{figure*}

In Fig. \ref{fig:HCN_vs_OH}, the OH absorption median velocities are plotted against the luminosities of the HCN-vib line, normalized to the total infrared luminosity of each galaxy. When available, the $J=3\text{--}2$ line has been chosen. In the four sources where both transitions have been observed, the luminosity ratio between the $J=4\text{--}3$ and $J=3\text{--}2$ lines lies in the range $2\text{--}3$, with a mean of $2.5$. For sources where only the $J=4\text{--}3$ line has been observed, we have therefore divided its luminosity by $2.5$ and included it in the plot with an extra $20\%$ uncertainty added in quadrature. The two galaxies with only emission in the $119$~$\mu$m doublet, I~Zw~1 and NGC~7469, are not included in the plot, but we note that neither of them have an HCN-vib detection.

In the plot in Fig. \ref{fig:HCN_vs_OH} we see that, in our sample, the sources with high $L_{\mathrm{HCN-vib}}/L_{\mathrm{IR}}$ ratios, in excess of $10^{-8}$, all lack fast outflows as traced by the median velocity of the OH $119$~$\mu$m doublet. However, galaxies without fast outflows do not necessarily have high $L_{\mathrm{HCN-vib}}/L_{\mathrm{IR}}$ ratios. The lack of sources in the upper left portion of the diagram is likely due to selection effects; that area should be populated by less obscured sources of lower luminosity that are not able to drive outflows or provide the necessary conditions for efficient HCN-vib excitation.

\section{Discussion}\label{sec:discussion}
Our comparison of HCN-vib line luminosities and OH $119$~$\mu$m absorption line velocities in nearby (U)LIRGs shows a trend of positive, or low negative, velocities, indicating inflows or slow outflows, in sources with bright HCN-vib emission relative to the total infrared luminosity. This might indicate that regions with the high mid-infrared brightness temperatures required for efficient population of HCN-vib are not affected by inflowing gas, but are destroyed by, or cannot form in the presence of, strong outflows. In this section, we discuss possible reasons for the lack of outflow signatures in the OH $119$~$\mu$m absorptions in HCN-vib luminous sources, and how the HCN-vib lines relate to other tracers of heavily obscured regions.

\subsection{Why do the HCN-vib luminous galaxies lack outflow signatures in the far-infrared OH lines?}
There are several scenarios that could explain the lack of outflow signatures in the galaxies that have bright HCN-vib emission. One is that there are no outflows, another that the outflows are hidden from detection using the median velocity of the OH $119$~$\mu$m absorption lines, and a third is that the distribution in Fig. \ref{fig:HCN_vs_OH} is due to orientation effects. These scenarios are discussed in the following sections. We also briefly discuss a possible evolutionary scenario in which the HCN-vib luminous galaxies are in a pre-feedback phase. Admittedly, our sample is quite heterogeneous in that it contains galaxies of different merger types, in different merger stages, of different optical classifications, and with infrared luminosities spanning more than an order of magnitude. It is therefore possible that there are several independent reasons for the lack of far infrared outflow signatures in HCN-vib luminous galaxies.

\subsubsection{No outflows}\label{sec:nooutflows}
If there are no outflows, it might be due to the HCN-vib luminous galaxies being intrinsically different from the galaxies with weaker HCN-vib emission. For example, \citet{vei13} find a tendency for systems with dominant AGN to have faster OH outflows; it might then be that the HCN-vib luminous systems, which do not show fast OH outflows, are instead starburst dominated. Indeed, in the two sources with the highest absolute HCN-vib luminosities, IRAS~17208-0014 and Arp~220, less than $10\%$ of the bolometric luminosities are estimated to come from AGN activity, based on the $15$ to $30$~$\mu$m continuum ratio \citep{vei13}. However, this diagnostic can be misleading in galaxies where the nuclei are optically thick into the far infrared. Furthermore, at least two other sources with strong HCN-vib emission, NGC~4418 and UGC~5101, have AGN fractions $\gtrsim 50\%$. While the uncertainty on the AGN fraction determined in this way is estimated to be $20\%$ on average, it is likely much higher in the strongly buried sources discussed here.

Another possible explanation for a real lack of outflows in some sources is that we are witnessing a pre-feedback phase in young systems that will later evolve into objects with weak HCN-vib emission and fast outflows. This possibility is discussed further in Sect. \ref{sec:evolution}. As discussed by \citet{gon17}, the lack of outflows could also be due to the fact that, in the most extremely buried sources, it is difficult to find paths with more moderate columns that can be efficiently accelerated to high velocities.

Interestingly, as \citet{aal15b} notes, fast molecular outflows have been found, with other tracers and at longer wavelengths, in some of the HCN-vib luminous systems. For example, a compact $v>800$~km\,s$^{-1}$ CO $J=2\text{--}1$ outflow has been found in the ULIRG IRAS~17208-0014 \citep{gar15} and an equally fast, compact, and collimated, HCN $J=1\text{--}0$ outflow has been detected in the ULIRG Arp~220 \citep{bar18}. Possible outflow signatures are also found in Zw~049.057 \citep{fal18} as well as in NGC~4418 and IRAS~22491-1808 \citep{flu18}, although at lower velocities. If we also consider the fact that UGC~5101 has faint high-velocity wings in its OH $119$~$\mu$m absorption lines \citep{vei13}, it turns out that all of the six most HCN-vib luminous sources have signatures of molecular outflows.

\subsubsection{Obscured, episodic, or collimated outflows}\label{sec:obscuredoutflows}
For the HCN-vib bright systems that do have outflows, there must be a reason why we do not see them in the far-infrared OH absorption lines. One possibility is that these objects have extreme optical depths in the far-infrared dust continuum, so that much of the outflowing gas cannot be seen due to obscuration by the dust. In at least some of the sources, high optical depths have indeed been inferred from radiative transfer modeling of \emph{Herschel} observations \citep[e.g.,][]{gon12,fal15}. If this is combined with young outflows that are still very compact, any signatures of them may be completely hidden in the far-infrared. In such a situation, inflows on larger scales, for example as the one seen in Arp~299A \citep{fal17}, may still be detectable. In fact, at least three of the five most obscured sources \citep[Arp~220, IRAS~17208-0014, and Zw~049.057;][]{baa89,fal18} have outflows that are seen in the OH lines at centimeter wavelengths but not in the far-infrared. In Zw 049.057, infrared and optical images reveal multiple clouds in a polar dust structure, which might suggest that the outflow in this galaxy is not steady in time (Gallagher et al. in prep.). It is thus possible that, in some galaxies, we are witnessing an episodic process where the nuclear region is alternating between feedback and accretion dominated states.

Looking at the spectra presented by \citet{vei13}, UGC~5101 and IRAS~17208-0014 even exhibit weak blueshifted line wings in the OH $119$~$\mu$m absorption lines. This indicates the presence of gas moving towards us, but as the absorptions at negative velocities are shallow, they do not have a strong effect on the median velocities of the lines. As discussed by \citet{gon17}, this may suggest that we are witnessing collimated outflows in these sources, as opposed to the wide-angle outflows that the OH $119$~$\mu$m lines seem to be primarily sensitive to. With all this in mind, it seems that the regions responsible for the bright HCN-vib emission are not destroyed by (fast) outflows in general, but rather by the kind of wide-angle outflows from the nuclear infrared-emitting region that are able to shift the median velocity of the OH $119$~$\mu$m ground state lines by several hundreds of km\,s$^{-1}$.

Turning to the two sources that show pure emission in the OH $119$~$\mu$m doublet, I~Zw~1 and NGC~7469, we note that they lack HCN-vib detections and that the upper limits to their $L_{\mathrm{HCN-vib}}/L_{\mathrm{IR}}$ ratios are among the lowest in our sample. Furthermore, NGC~7469 has also been observed in the OH $65$~$\mu$m doublet, another tracer of obscured nuclei, but no absorption was detected \citep{gon15}. This is consistent with the results of \citet{vei13} that the sources with pure OH emission are the ones where an AGN dominates the luminosity and that these objects represent a phase where the feedback has subsided after clearing a path through the dusty surroundings.

\subsubsection{Orientation effects}\label{sec:orientation}
An alternative explanation to the distribution of galaxies in Fig. \ref{fig:HCN_vs_OH} is that it is produced by a simple orientation effect. For example, using the OH $119$~$\mu$m doublet, \citet{vei13} found unambiguous evidence of outflows in $70\%$ of their sources, a detection rate which is consistent with all sources having molecular outflows, with an average opening angle of ${\sim}145\degr$. The sources with no outflow signatures would then be those that are oriented approximately edge-on, with outflows perpendicular to our line of sight. It is harder to explain why these would also be the sources with bright HCN-vib emission, as the obscuration in a dusty edge-on disk would rather have the effect of lowering the visible HCN-vib flux, unless the line is masing as seen towards some circumstellar envelopes \citep[e.g.,][]{luc89,bie01,men18}.

However, the comparison so far has been based on the HCN-vib luminosity divided by the total infrared luminosity which, due to the higher dust opacity in the infrared, is more affected by orientation effects. The infrared luminosity that we see will thus depend on the orientation, with higher values for face-on galaxies. This effect may indeed be significant; for example, \citet{efs14} estimated that the true infrared luminosity of IRAS 08572+3915 may be as high as $1.1 \times 10^{13}$~L$_{\odot}$, more than five times higher than the value inferred assuming the luminosity is isotropic. Another possible example is Arp~220 where the total infrared luminosity estimated from the IRAS fluxes is ${\sim}1.7\times10^{12}$~L$_{\odot}$ while \citet{sak17} estimate the bolometric luminosity from the western nucleus alone to be ${\sim}3\times10^{12}$~L$_{\odot}$.

If we do not divide $L_{\mathrm{HCN-vib}}$ by the total infrared luminosity, we see that Mrk~231 has an HCN-vib luminosity comparable to those of the HCN-vib bright galaxies. However, it is still less luminous than in the ULIRGs Arp~220~W and IRAS~17208-0014, and only a factor of two or three brighter than in the LIRGs Zw~049.057, IC~860, and NGC~4418. Furthermore, none of the other sources with $v_{50}\lesssim-200$~km\,s$^{-1}$ have strong HCN-vib emission. Clearly, a larger sample will be required to investigate this issue further. If the distribution of sources in Fig. \ref{fig:HCN_vs_OH} is indeed due to an orientation effect, more sources with both strong HCN-vib emission and outflow signatures in the far-IR OH lines should show up in an extended sample.

\subsubsection{Evolution}\label{sec:evolution}
\begin{figure}
   \centering
   \includegraphics[width=0.5\textwidth]{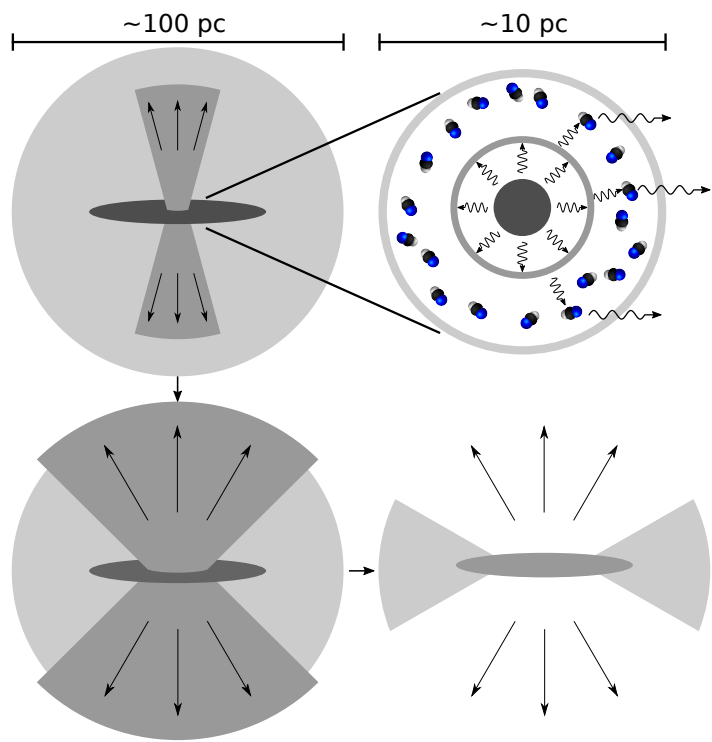}
   \caption{Schematic view of a possible evolutionary scenario. The upper left panel shows the most obscured phase where both the mid-infrared core and the outflow are still completely embedded in a layer of cooler dust. The upper right panel shows the conditions in the mid-infrared core in this phase. A central energy source is heating the dusty core through radiative trapping. Mid-infrared photons, which cannot escape the dusty environment, vibrationally excite the HCN molecules which may then undergo rotational transitions and emit (sub)millimeter photons that are able to penetrate the obscuring dust. In the lower left panel, the outflow has broken through the cooler layer of dust and widened, making it visible in the far-infrared. The lower right panel shows a third stage where the outflow has cleared a path through the surrounding dust layer. In the two final stages, conditions in the core are no longer sufficient for luminous HCN-vib emission to occur.}
              \label{fig:scenario}%
\end{figure}

Following \citet{aal15b}, we present a possible evolutionary scenario in which bright HCN-vib emission is tracing extremely obscured nuclei in a phase that occurs before the nuclear feedback is able to drive wide-angle outflows, or at least before such outflows can be observed in the far-infrared OH lines. In this scenario, the deeply embedded regions which provide the necessary conditions for HCN-vib excitation can coexist with fast collimated outflows, but are disrupted once nuclear wide-angle outflows have developed. A schematic view of one possible version of the scenario is presented in Fig. \ref{fig:scenario}. Here, the most obscured sources consist of a mid-infrared core, responsible for the HCN-vib emission, surrounded by an obscuring layer of cooler dust that accounts for the far-infrared emission. Outflows in these objects are collimated and often so compact that they are still embedded in the outer layer of dust, and thus obscured from view in the far-infrared. In the next phase, the outflows have broken through the cooler dust layer and widened considerably, making them clearly detectable in most galaxies. As gas and dust is transported from the regions around the nucleus, the conditions in the mid-infrared core change, becoming less favorable for the excitation of HCN-vib. Finally, when a path has been cleared through the surrounding dust, OH is no longer detectable in absorption and the conditions for HCN-vib excitation in the core are no longer met.

A similar evolutionary sequence has been suggested by \citet{gon17b}, and we note that, apart from UGC~5101, the HCN-vib luminous sources from our sample coincide with the sources in their extremely buried pre-feedback phase. It should be noted that, even if an evolutionary scenario like this applies in some cases, all sources in our sample may not be on the same evolutionary track. For example, a lot of nuclear power is required in order to push large columns of gas to high velocities, and LIRGs like NGC~4418 and Zw~049.057 may therefore never develop powerful wide-angle outflows detectable in the far-IR OH lines. Instead, they might evolve along the horizontal axis in Fig. \ref{fig:HCN_vs_OH}.

Finally, an interesting comparison can be made with the highly collimated molecular outflows that are found around low-mass protostars in the earliest (class 0) stages of their formation \citep[e.g.,][]{kon00,cod14}. With time, gas at larger and larger angles from the outflow axis may be entrained, eventually sweeping away most of the envelope around the star \citep[e.g.,][]{arc06}. Furthermore, vibrationally excited HCN has been detected toward molecular hot cores around massive protostars using the direct $l$-type transitions, that is transitions between the $l$-type levels inside a rotational state, at cm-wavelengths \citep[e.g.,][]{rol11a}.

\subsection{HCN-vib compared to other tracers of obscured nuclear regions.}
As discussed in \citet{aal15b}, a prerequisite for the excitation of HCN-vib is the presence of warm and dusty regions, making it a good tracer of deeply buried nuclei. Following this logic, our results indicate that the most obscured nuclei lack evolved, wide-angle outflows. Interestingly, the results of \citet{gon17}, who use the equivalent width of the OH $65$~$\mu$m doublet as a measure of the obscuration, suggest that the fastest outflows arise in some of the most obscured nuclei. They do however point out that sources with high equivalent width in OH $65$~$\mu$m do not necessarily have fast outflows. By comparing our Fig. \ref{fig:HCN_vs_OH} with their Fig. 4, we see that five of the six most HCN-vib luminous sources that are included in both samples also belong to the group of sources with high equivalent width but low outflow velocity, indicating that they are obscured also when using the OH $65$~$\mu$m equivalent width as a measure. On the other hand, five of the sources with low HCN-vib luminosity are also found among the sources with high OH $65$~$\mu$m equivalent width, indicating that they also contain obscured nuclei. It appears that the two measures trace different parts or physics of the galaxy nuclei. Indeed, \citet{gon17} state that high equivalent widths require dust temperatures in excess of $60$~K, while dust temperatures above $100$~K are required for bright HCN-vib emission \citep{aal15b}. It is thus possible that bright HCN-vib emission exclusively traces more obscured parts of the nuclei, where the dust is hotter.

In a similar manner, \citet{spo13} also find that the fastest outflows are found in those sources that are still deeply embedded as indicated by strong mid-infrared silicate absorptions. However, as already noted by \citet{aal15b}, \citet{gon15} find that silicate absorptions are biased toward relatively unobscured mid-infrared emitting regions, again indicating that the HCN-vib is actually tracing a more extreme form of obscuration.  

\section{Conclusions}\label{sec:conclusions}
We explore a possible correlation between the strength of molecular outflows in the far-infrared OH $119$~$\mu$m lines and the luminosity of rotational lines of vibrationally excited HCN. A simple comparison of HCN-vib line luminosities, normalized to the total infrared luminosity of the host galaxy, and the median velocities of OH $119$~$\mu$m absorption lines shows that galaxies with unusually bright HCN-vib emission tend to lack fast molecular outflows, but that galaxies without fast outflows do not necessarily have bright HCN-vib emission. This may be an orientation effect or something that reflects a true difference between the sources, for example that the most obscured sources cannot drive wide-angle outflows or that their outflows are young and have not yet dispersed the nuclear gas and dust concentrations. Following \citet{aal15b}, we present an evolutionary sequence, that may apply to some of the sources, in which bright HCN-vib emission is tracing extremely obscured nuclei in a phase that occurs before the onset of wide-angle outflows. Once these massive outflows have been launched, they quickly disrupt the deeply embedded regions responsible for the bright HCN-vib emission. We note, however, that more studies are needed in order to extend the sample and explore the relation further. As the \emph{Herschel} Space Observatory, which was used for the OH outflow measurements, is no longer operational, such studies should concentrate on observations of HCN-vib in galaxies which have already been observed in the OH $119$~$\mu$m doublet. Another possibility is to search for hidden outflows using the radio lines of OH, either in its ground state \citep{baa89} or in rotationally excited states \citep{fal18}.

\begin{acknowledgements}
This paper makes use of the following ALMA data: ADS/JAO.ALMA\#2015.1.00708.S, ADS/JAO.ALMA\#2016.1.00140.S. ALMA is a partnership of ESO (representing its member states), NSF (USA) and NINS (Japan), together with NRC (Canada), MOST and ASIAA (Taiwan), and KASI (Republic of Korea), in cooperation with the Republic of Chile. The Joint ALMA Observatory is operated by ESO, AUI/NRAO and NAOJ.

This research has made use of the NASA/IPAC Extragalactic Database (NED), which is operated by the Jet Propulsion Laboratory, California Institute of Technology, under contract with the National Aeronautics and Space Administration. 
   \end{acknowledgements}
\bibliographystyle{bibtex/aa} 
\bibliography{ref} 

\begin{appendix}

  \section{Vibrationally excited HCN in IRAS 15250+3609}\label{app:15250}
  The ULIRG IRAS 15250+3609 was observed in the HCO$^{+}$ $J=3\text{--}2$ line by \citet{ima16a}.  They interpreted a peak close to the expected position of the HCN-vib line as an outflow signature in the HCO$^{+}$ line. Their interpretation is supported by the fact that similar features are seen in the HCN $J=3\text{--}2$ and $J=4\text{--}3$ lines \citep{ima16a,ima18}. However, relative to the main lines, the sub-peak close to the HCO$^{+}$ line is stronger than the one close to the HCN line. This could be due to differing abundances or excitation in the outflowing gas but it could also be a contribution from HCN-vib. Furthermore, IRAS~15250+3609 has a high equivalent width in the OH $65$~$\mu$m doublet \citep{gon15}, indicating that the galaxy is highly obscured, as well as an HCN absorption at $14$~$\mu$m \citep{lah07}. Here, we have attributed the feature as completely due to HCN-vib emission, but it is likely that it is indeed blended with an HCO$^{+}$ outflow signature and the value given here should be considered an upper limit to the HCN-vib luminosity. 

\section{Previously unpublished HCN-vib detections}\label{app:HCN}
We found two sources with previously unpublished detections of HCN-vib in the ALMA science archive, IRAS~12224-0624 (project 2015.1.00708.S, PI: L. Armus), and IRAS~05189-2524 (project 2016.1.00140.S, PI: D. Iono). The observations of IRAS~05189-2524 were conducted during two runs on 2016 November 12 and 15 with a total of $42$ antennas with baseline lengths ranging between $15.1$~m and $1.0$~km. A spectral window centered at $341.8$~GHz covered a bandwidth of $1.875$~GHz (${\sim}1650$~km\,s$^{-1}$ at the frequency of HCN-vib $J=4\text{--}3$), with a frequency resolution of $3.9$~MHz. During both runs \object{J0522-3627} was used as bandpass and flux calibrator, and \object{J0457-2324} was used as phase calibrator. The total on-source time of the observations was $3340$~s and the final sensitivity achieved was $0.5$~mJy\,beam$^{-1}$ per $20$~km\,s$^{-1}$ (${\sim}24$~MHz) channel. For the imaging we used Briggs weighting with a robustness factor of 0.5. The resulting beam size is $0.29\arcsec\,\times\,0.22\arcsec$ (PA ${\sim}63\degr$).

The observations of IRAS~12224-0624 were conducted during a single run on 2016 April 27 with a total of $38$ antennas with baseline lengths ranging between $15.1$~m and $452.8$~m. A spectral window centered at $347.4$~GHz covered a bandwidth of $1.875$~GHz (${\sim}1600$~km\,s$^{-1}$ at the frequency of HCN-vib $J=4\text{--}3$), with a frequency resolution of $3.9$~MHz. The calibrators used were \object{J1256-0547}, which was used for bandpass and flux calibration, and \object{J1246-0730}, which was used for phase calibration. The total on-source time of the observations was $395$~s and the final sensitivity achieved was $1.5$~mJy\,beam$^{-1}$ per $20$~km\,s$^{-1}$ (${\sim}24$~MHz) channel. For the imaging we used Briggs weighting with a robustness factor of 0.5. The resulting beam size is $0.51\arcsec\,\times\,0.44\arcsec$ (PA ${\sim}82\degr$).

\begin{figure}[ht]
   \centering
   \includegraphics{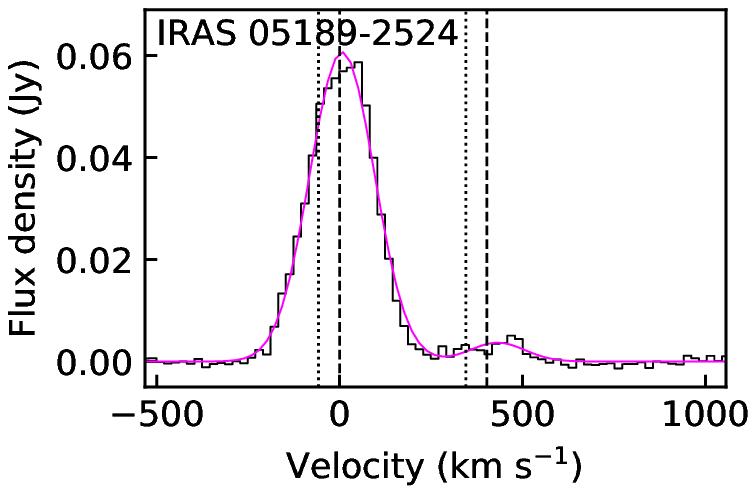}
   \includegraphics{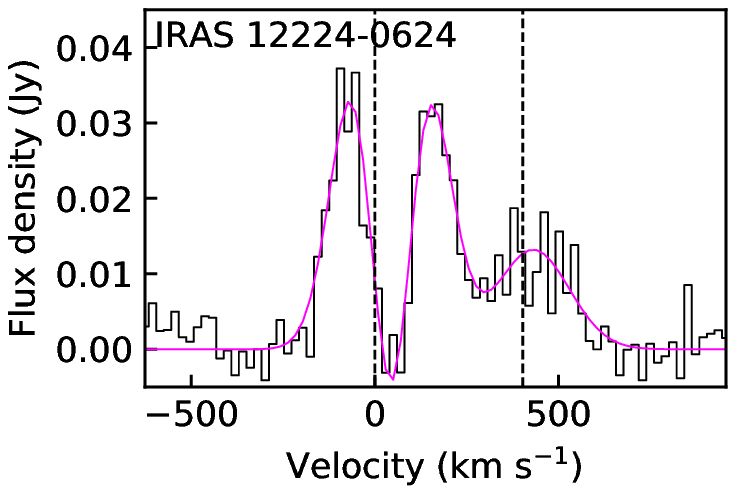}
   \caption{Spectral fits to the HCO$^{+}$ and HCN-vib $J=4\text{--}3$ lines in the two galaxies IRAS~05189-2524 and IRAS~12224-0624. The solid black histograms represent the data and the solid magenta lines are the best fits to the data. The velocity scale is set relative to the frequency of the HCO$^{+}$ $J=4\text{--}3$ line. Dashed vertical lines indicate the expected positions of the HCO$^{+}$ and HCN-vib $J=4\text{--}3$ lines given the adopted redshifts. The dotted lines in the plot for IRAS~05189-2524 indicate the expected positions given the redshift adopted by \citet{vei13}.}
              \label{fig:new_hcnvib}%
  \end{figure}

For both sources, the Common Astronomy Software Applications \citep[CASA;][]{mcm07} package was used to reapply the calibration and perform imaging. The line fluxes used to calculate the HCN-vib luminosities were then found by fitting Gaussians to the HCO$^{+}$ and HCN-vib lines in spectra that were spatially integrated over the sources. These spectra are presented in Fig. \ref{fig:new_hcnvib}. We note that the wavelengths of the HCO$^{+}$ and HCN-vib lines in IRAS~05189-2524 are offset from those expected when using the redshift of $0.0426$ adopted by \citet{vei13}. They are however consistent with the redshift of ${\sim}0.0428$ found from the CO $J=1\text{--}0$ observations reported by \citet{san91}. Using this redshift instead of the one adopted by \citet{vei13} to determine the median OH absorption velocity results in an extra velocity shift of $-45$~km\,s$^{-1}$.

  \section{New OH outflow measurements}\label{app:OH}
\begin{figure}[ht]
   \centering
   \includegraphics{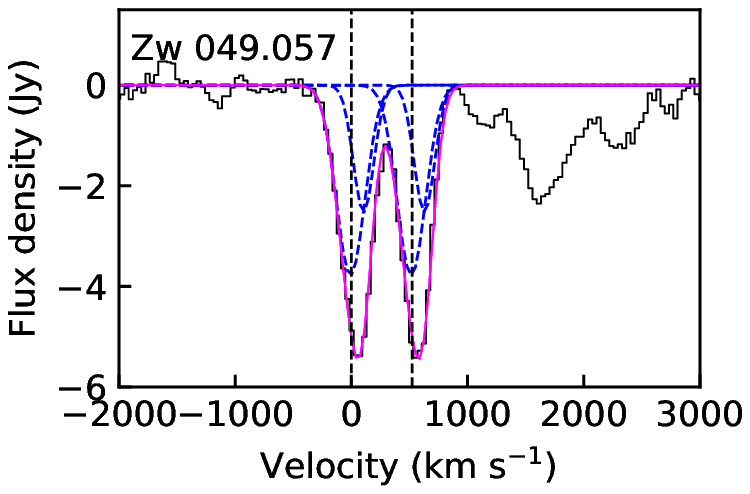}
   \includegraphics{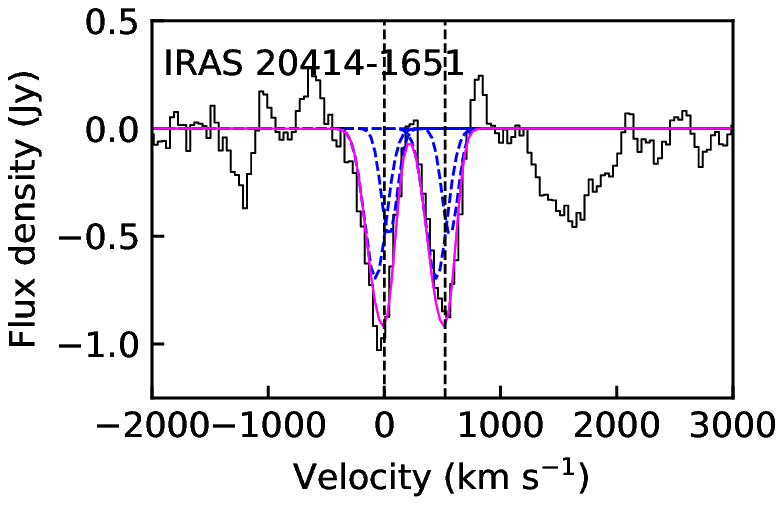}
   \includegraphics{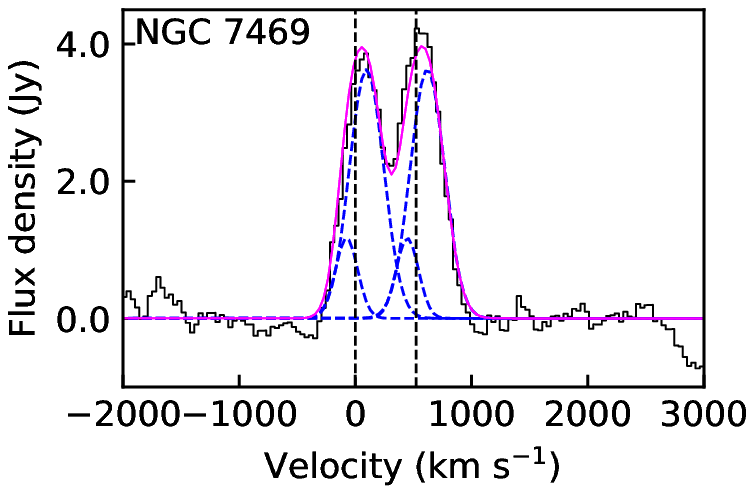}

   \caption{Spectral fits to the OH $119$~$\mu$m absorption lines in the galaxies Zw~049.057, IRAS~20414-1651, and NGC~7469. The solid black histograms represent the data, the solid magenta lines are the best multi-component fits to the data, and the dashed blue lines are the individual components. The velocity scale is set relative to the frequency of the blue component of the doublet. Dashed vertical lines indicate the expected positions of the two absorption components given the adopted redshifts.}
              \label{fig:new_oh}%
  \end{figure}

  Three of the sources with existing HCN-vib observations also had observations of the OH doublet at $119$~$\mu$m taken with the Photodetector Array Camera and Spectrometer \citep[PACS;][]{pog10} on \emph{Herschel} but were not included in the sample of \citet{vei13}. The observations of Zw~049.057 (OBSID: 1342248368, PI: E. Gonz\'alez-Alfonso) were conducted on 2012 July 20 for a duration of $1478$~s, the observations of IRAS~20414-1651 (OBSID: 1342217908, PI: D. Farrah) were conducted on 2011 April 5 for a duration of $8879$~s, and the observations of NGC~7469 (OBSID: 1342235840, PI: E. Gonz\'alez-Alfonso) were conducted on 2011 December 31 for a duration of $1086$~s. All observations were performed in high spectral sampling, range spectroscopy mode. The observations had been processed with version 14.2 of the standard pipeline and were not in need of reprocessing. In all sources, the nuclear far-IR emission is spatially unresolved in the central $9.4$\arcsec\ (${\sim}2$~kpc) spaxel of the PACS $5$~x~$5$ spaxel array. As the central PACS spaxel is smaller than the point spread function of the spectrometer, the spectrum was extracted using the point source correction task in the \emph{Herschel} interactive processing environment \citep[HIPE;][]{ott10} version 14.0.1. Before analyzing the absorption lines, polynomials of order two were fitted to the continuum and then subtracted from the spectra. For consistency, the profiles of the $119$~$\mu$m OH doublets were modeled using the same procedure as in \citet{vei13}. Each line was fitted with two Gaussian components characterized by their amplitude, position, and width. The separation between the two lines of the doublet was fixed at $0.208$~$\mu$m in the rest frame and the amplitude and width of the two lines were the same for each component. The median velocities of the absorptions ($v_{\mathrm{50}}$(abs)) were then determined from these fits. The fitting procedure was carried out using the spectroscopic analysis toolkit \emph{PySpecKit} \citep{gin11} and the continuum subtracted spectra with the fits overplotted are presented in Fig. \ref{fig:new_oh}. 

\end{appendix}

\end{document}